\documentclass[a4paper,11pt]{article}
\usepackage[latin1]{inputenc}
\usepackage[colorlinks=true,ps2pdf=true]{hyperref}
\usepackage[dvips]{graphics}
\usepackage{psfrag}
\usepackage{amsmath}
\usepackage{amssymb}
\usepackage[english]{babel}
\usepackage{url}

\newcommand{\matching}[1]{M(#1)}
\newcommand{\maxmatching}[1]{M^*(#1)}
\newcommand{\incident}[1]{X(#1)}
\newcommand{\maxdegree}{\Delta}
\newcommand{\email}{\url}

\makeatletter

\long\def\@doifundefined#1#2{\def\reserved@jhh{\relax}%
 \expandafter\ifx\csname#1\endcsname\relax
    \def\reserved@jhh{#2}%
 \fi
 \reserved@jhh}
\newcommand{\provideenvironment}[2]
  {\@doifundefined{#1}{\newenvironment{#1}#2}}
\newcommand{\providetheorem}[2]
  {\@doifundefined{#1}{\newtheorem{#1}#2}}

\def\rcs$#1${#1}
\newcommand{\version}[1]{\thanks{\rcs#1}}
\newcommand{\neighb}[2][{}]{\Gamma_{#1}({#2})}	
\newcommand{\@Setstar}[1]{\left\{{#1}\right\}}
\newcommand{\@Set}[2]{\@Setstar{{#1},\ldots,{#2}}}
\newcommand{\Set}{\@ifstar{\@Setstar}{\@Set}}	

\newcommand{\Qbind}[2]{#1\mskip1mu #2}
\newcommand{\Quant}[3]{\left(#1
  \if\relax#2\relax :: 
  \else : #2 :: 
  \fi 
  #3\right)}
\newcommand{\Qall}[1]{\Qbind{\forall}{#1}}
\newcommand{\all}[3]{\Quant{\Qall{#1}}{#2}{#3}}	

\newcommand{\call}[2]{\id{#1}(#2)}
\newcommand{\id}[1]{\ensuremath{\mathit{#1}}}	
\newcommand{\etal}{\textit{et al.}~}	
\newcommand{\term}[1]{\emph{#1}}		
\newcommand{\ie}{{i.e.},\ }		
\newcommand{\lineno}[1]{\ensuremath{\mathrm{#1}}}
\newcommand{\keyw}[1]{\ensuremath{\mathbf{#1}}} 

\newcommand{\assign}{\mathrel{:=}}	
\newcommand{\IF}{\keyw{if}~}
\newcommand{\WHILE}{\keyw{while}~}
\newcommand{\DO}{\keyw{do}~}
\newcommand{\TO}{\keyw{to}~}
\newcommand{\FORALL}{\keyw{forall}~}
\newcommand{\RETURN}{\keyw{return}~}

\usepackage{ifthen}
\usepackage{float}
\usepackage{theorem}
\floatstyle{plain}
\newfloat{protocol}{t}{lop}[section]
\floatname{protocol}{Protocol}
\newcounter{JHH@proglinei} 
\renewcommand{\theJHH@proglinei}
	{\arabic{JHH@proglinei}}
\newcounter{JHH@proglineii}[JHH@proglinei] 
\renewcommand{\theJHH@proglineii}
	{\theJHH@proglinei.\arabic{JHH@proglineii}}
\newif\ifJHH@docr	 
\newcommand{\JHH@cr}{\ifJHH@docr\\ \else\global\JHH@docrtrue\fi}
\newcommand{\nl}[1][-]{\JHH@cr %
  \refstepcounter{JHH@proglinei}%
  \ifthenelse{\equal{#1}{-}}{}{\label{ln-#1}}%
  \lineno{\theJHH@proglinei}\>}
\newcommand{\nnl}{\@ifstar{\stepcounter{JHH@proglinei}\nnldo}{\nnldo}}
\newcommand{\nnldo}[1][-]{\JHH@cr %
  \refstepcounter{JHH@proglineii}%
  \ifthenelse{\equal{#1}{-}}{}{\label{ln-#1}}%
  \lineno{\theJHH@proglineii}\>}
\newcommand{\ul}{\JHH@cr\@ifstar{\ignorespaces}{\>}}
\newcommand{\com}{\@ifstar{\@comfl}{\@com}}
\newcommand{\@com}[1]{$(*$ #1 $*)$}
\newcommand{\@comfl}[1]{\`\@com{#1}}
\newenvironment{program}[1][1]%
{\begin{tabbing}%
 \setlength{\tabbingsep}{0pt}%
 \ifthenelse{#1 = 0}{}{\setcounter{JHH@proglinei}{#1}%
	\addtocounter{JHH@proglinei}{-1}}%
 \lineno{9999}\=\lineno{9999}\=\lineno{9999}\=\kill%
 \global\JHH@docrfalse\ignorespaces}%
{\end{tabbing}}
\providecommand{\qed}{}
\newcommand{\jhhproofend}{$\triangleleft$}
\provideenvironment{proof}%
  {{\noindent\hbox{\bf Proof: }\ignorespaces}%
  {\unskip\penalty10000\hglue1cm\penalty10000\hfill\jhhproofend
   \par\vspace{\baselineskip}}}
\theoremstyle{plain}
\theorembodyfont{\it}
\theoremheaderfont{\bf}
\providetheorem{theorem}{{Theorem}[section]}
\providetheorem{property}{[theorem]{Property}}
\providetheorem{claim}{[theorem]{Claim}}
\providetheorem{proposition}{[theorem]{Proposition}}
\providetheorem{lemma}{[theorem]{Lemma}}
\providetheorem{corollary}{[theorem]{Corollary}}
\providetheorem{definition}{[theorem]{Definition}}
\providetheorem{assumption}{[theorem]{Assumption}}
\providetheorem{observation}{[theorem]{Observation}}
\providetheorem{axiom}{[theorem]{Axiom}}
\providetheorem{property}{[theorem]{Property}}
\makeatother

\newcommand{\SEND}{\keyw{send}~}
\newcommand{\RECEIVE}{\keyw{receive}~}
\newcommand{\THEN}{$\;\longrightarrow\;$}

\newcommand{\FROM}{\keyw{from}~}
\newcommand{\msg}[1]{\langle \textit{#1} \rangle}

\title{Simple Distributed Weighted Matchings\version{$Id: weighted-matchings.tex,v 1.3 2004/10/19 08:30:12 jhh Exp $}}

\author{Jaap-Henk Hoepman\\
Nijmegen Institute for Computing and Information Sciences (NIII)\\
  Radboud University Nijmegen\\
  P.O. Box 9010, 6500 GL \ Nijmegen, 
  the Netherlands\\ \email{jhh@cs.ru.nl}}

\begin{document}

\maketitle

\bibliographystyle{alphacm-}

\begin{abstract}
Wattenhofer~\etal\cite{wattenhofer2004weightedmatching} derive a complicated
distributed algorithm to compute a weighted matching of an arbitrary weighted
graph, that is at most a factor $5$ away from the maximum weighted matching of
that graph. We show that a variant of the obvious sequential greedy
algorithm~\cite{preis1999maxmatching}, 
that computes a weighted matching at most a factor $2$ away from the maximum,
is easily distributed. This yields the best known distributed approximation
algorithm for this problem so far.
\end{abstract}

\section{Introduction}

A \term{matching} $\matching{G}$ of a graph $G=(V,E)$ is any subgraph of $G$
where no two edges are incident to the same vertex. Let $w(e)$ be the weight of
an edge $e \in E$ of $G$, where $w(e) > 0$. Define the weight $w(G)$ of a graph
$G$ to be the sum of the weights of all its edges. Then a
\emph{maximum weighted matching} $\maxmatching{G}$ of $G$ is a matching whose
weight is the maximum among all matchings of $G$.

Surprisingly, few distributed algorithms to compute (an approximation of) the
maximum weighted matching of the communication graph are known. For unweighted
graphs, there are deterministic distributed algorithms computing the
\emph{maximal} matching in trees~\cite{karaata2000selfstabmatching}, and
bipartite and general graphs~\cite{chattopadhyay2002selfstabmatching}. 
Randomised algorithms for the 
general case~\cite{israeli1986maxmatching} also exist. 

For weighted graphs, Uehara~\etal\cite{uehara2000maxmatching} present a
constant time 
distributed algorithm that computes a weighted matching that is $O(\maxdegree)$
away from the maximum (where $\maxdegree$ is the maximum degree of the graph).
Recently,
Wattenhofer~\etal\cite{wattenhofer2004weightedmatching} derived a complicated
randomised distributed algorithm to compute a weighted matching $\matching{G}$
with \term{approximation ratio} $5$, \ie such that
$w(\matching{G}) > \frac{1}{5} w(\maxmatching{G})$.

For sequential algorithms, the problem is much better studied.
For unweighted graphs, Micali and Vazirani~\cite{micali1980maxmatching} 
present an $O(\sqrt{|V|} |E|)$ time algorithm that computes a maximal matching.
For weighted graphs Gabow~\cite{gabow1990weightedmatching} gives an 
$O(|V| |E| + |V|^2 \log |V| )$ time algorithm, computing the maximum weighted
matching. Both return an exact solution, and not approximations.

Recently, there have improvements in the performance of sequential
al\-go\-rithms to approximate the maximum weighted matching of a graph, that
require much less running time than the exact algorithms.

The obvious greedy sequential algorithm (that each time adds the remaining
heaviest edge) computes a weighted matching 
at most a factor $2$ away from the maximum, in running time 
$O(|E| \log |V|)$~\cite{avis1983weightedmatching}. 
Preis~\cite{preis1999maxmatching} showed that selecting locally heaviest edges
instead of globally heavy edges achieves the same approximation, improving the
running time to $O(|E|)$.
Using a path-growing algorithm, Drake~\etal\cite{drake2003weightedmatching}
achieve the same running time and performance ratio. 

Later, Drake~\etal\cite{drake2003maxmatching2} improved the approximation 
to $3/2+\epsilon$, using a slowly converging algorithm using the concept of
augmenting paths. Pettie~\etal\cite{pettie2004maxmatching} present both a
deterministic and a randomised algorithm achieving the same approximation in
running time $O(|E| \log\frac{1}{\epsilon})$.

In this paper, we show that Preis's algorithm is easily distributed
deterministically. This gives us an $O(|E|)$ time deterministic distributed
algorithm that computes a weighted matching with an approximation ratio $2$,
the best known so far.

\section{A distributed greedy algorithm}

\begin{protocol}
\begin{center}
\begin{program}
\ul $\matching{G} = \emptyset$
\ul \WHILE $E \neq \emptyset$
\ul \DO \= pick locally heaviest edge $e$ from $E$
\ul     \> add $e$ to $\matching{G}$
\ul     \> remove $e$ and all edges incident to $e$ from $E$     
\ul \RETURN $\matching{G}$
\end{program}
\end{center}
\caption{Sequential greedy weighted matching protocol.}
\label{prot-seq}
\end{protocol}

\noindent
We derive a distributed variant from the sequential protocol~\ref{prot-seq} 
due to Preis~\cite{preis1999maxmatching}, who proved that this protocol
approximates the maximum matching by a factor $2$. 
\begin{lemma}[Preis]
\label{lem-seqapprox}
Protocol~\ref{prot-seq} returns for any graph $G$ a matching $\matching{G}$ such
that $w(\matching{G}) \ge \frac{1}{2} w(\maxmatching{G})$.
\end{lemma}
We assume an asynchronous distributed system where nodes in $V$ can send
messages to their neighbours over the communication links $E$. We set
$G=(V,E)$. Message passing is asynchronous but reliable. 

Let $\neighb{v}$ be the set of neighbours of $v$ in $G$. Define
\[
    \call{candidate}{u,N} = v \in N 
	\text{~s.t. $\all{v' \in N}{}{w(u,v) \ge w(u,v')}$}
\]
to be the node in the set of remaining neighbours reached by the locally
heaviest edge as seen from $u$. 

In the distributed version of the greedy protocol (see
protocol~\ref{prot-dist}), each node $u$ start with a set $N$ equal to all its
neighbours in the graph. A node sends a \term{request} to its current
\term{candidate} neighbour connected to it over the locally heaviest edge (from
$u$'s point of view). This request is either granted (because the neighbour
replies with a request to $u$ as well, meaning that both see this as the
locally heaviest edge), or the edge is eventually dropped (if the target node
added a different edge to the matching, dropping all remaining edges from the
graph).  The set $N$ maintains the set of neighbours that are still reachable by
non-dropped edges. The set $R$ contains all nodes from which requests have been
received.
If an edge over which a request was sent is dropped, $u$
sends a new request to a new candidate in $N$. 

\begin{protocol}
\begin{center}
\begin{program}
\ul $R \assign \emptyset$
\ul $N \assign \neighb{v}$
\ul $c \assign \call{candidate}{v,N}$
\ul \IF $c \neq \bot$ \THEN \SEND $\msg{req}$ \TO $c$
\ul \WHILE $N \neq \emptyset$
\ul \DO \= \RECEIVE $m$ \FROM $u$
\ul     \> \IF $m = \msg{req}$ 
              \THEN \= $R \assign R \cup \Set*{u}$
\ul     \> \IF $m = \msg{drop}$
              \THEN \= $N \assign N \setminus \Set*{u}$
\ul     \>          \> \IF $u = c$ 
                       \THEN \= $c \assign \call{candidate}{v,N}$
\ul     \>          \>       \> \IF $c \neq \bot$ 
                                \THEN \SEND $\msg{req}$ \TO $c$
\ul     \> \IF $c \neq \bot \wedge c \in R$ 
           \THEN \= \FORALL $w \in N \setminus \Set*{c}$ \SEND $\msg{drop}$ \TO $w$
\ul     \>       \> $N \assign \emptyset$
\ul \{ if $c \neq \bot$ then $(v,c) \in M$ \}
\end{program}
\end{center}
\caption{Distributed greedy weighted matching protocol (node $v$).}
\label{prot-dist}
\end{protocol}

\subsection{Proof of correctness}

In the proof of the protocol we assume all edge weights are unique. If they
aren't, node identities can be added to break symmetry. For node $v$ we write
$N_v$ and $c_v$ for its local variables.

The main idea of the proof is
to show that protocol~\ref{prot-dist} essentially simulates
protocol~\ref{prot-seq}. Define for a run of protocol~\ref{prot-dist}
the event that two nodes $u,v$ \emph{match} when $u$ receives from $v$ a
$\msg{req}$ message while $u$ sent a $\msg{req}$ message to $v$ before 
(\ie $c_u = v$ and $c_v = u$). 

Consider a run of protocol~\ref{prot-dist} on input $G$.
Consider all matching events $x_i$ in that run (as defined
above), and order them in order of occurrence ($x_1$ being the first, $x_0$
is the wake up event of the algorithm). 
Let matching event $x_i$ match the pair $(u_i,v_i)$ (which adds edge
$e_i = (u_i,v_i)$ to the matching). 
Define for event $x_i$ the set of remaining edges $E_i$
inductively as follows. Set $E_0 = E$, and set 
\[
	E_i = E_{i-1} \setminus \Set*{\text{all edges incident to $u_i$ and $v_i$}}~.
\]

\begin{proposition}
\label{prop-onemess}
In protocol~\ref{prot-dist}, each node sends at most one message over each
incident edge.
\end{proposition}
\begin{proof}
A node only sends a $\msg{req}$ message after it removed the previous
candidate from the $N$. It sends a $\msg{drop}$ message to all remaining nodes
in $N$ (to which it didn't send a $\msg{req}$ yet), except for the
current candidate, and then terminates by setting $N=\emptyset$.
\end{proof}

\begin{proposition}
\label{prop-N-E}
After $x_i$, and before $x_{i+1}$ (if it occurs), if
$(u,v) \in E_i$ then $u \in N_v \land v \in N_u$.
\end{proposition}
\begin{proof}
The proposition holds initially. Consider the moment when a node $u$ is
removed from a set $N_v$. This either happens when $v$ receives a matching
$\msg{req}$ by some node $w = c_v$, or a $\msg{drop}$ from $u$.
In the first case, a match event $x_i$ occurs and all edges incident to
$v$ are removed from $E_{i-1}$ to construct $E_i$, including $(u,v)$. In the
second case, if $u$ sent a $\msg{drop}$ message, it was because of another
match event $x_j$ equal or before $x_i$ in which all edges incident to $u$ were
removed (similar to the first case). As $E_j \supseteq E_i$, the proposition
follows.  
\qed
\end{proof}

\begin{proposition}
\label{prop-c-E}
For all $i$, we have $e_i \in E_{i-1}$.
\end{proposition}
\begin{proof}
Suppose not. If $e_i=(u,v)$ is removed from $E_{j-1}$ (to construct $E_j$) 
for some $j<i$, then
a matching event $(u,w)$ (or $(v,w)$) occurred removing all edges
incident to $u$. But then $c_u=w$ remains forever, contradicting that $u$ is
involved in matching event $x_i$ (even if $w=v$). 
\end{proof}

\begin{proposition}
\label{prop-term}
Protocol~\ref{prot-dist} terminates for every node in the graph, 
with $E_t = \emptyset$ for some $t$.
\end{proposition}
\begin{proof}
By proposition~\ref{prop-onemess}, a node can receive at most one 
$\msg{req}$ from each neighbour. After all those are received, 
each iteration of the loop removes elements from $N_v$. Hence, eventually
$N_v = \emptyset$ and $v$ terminates, unless
$v$ waits for receipt of a message forever in the first line of the
loop. But then $N_v \neq \emptyset$ and hence 
$c_v = u \neq \bot$ for some $u$. This means a $\msg{req}$ message was sent
to $u$. Then either $v \in N_u$, or a $\msg{drop}$ message is in transit to
$v$ (contradicting that $v$ waits forever for a new message).
But if $v \in N_u$ it will either become a candidate for $u$ (in which case
$u$ sends $\msg{req}$ to $v$), or $u$ finds another candidate, sending a
$\msg{drop}$ to all remaining nodes in $N_u$ including $v$.

To show that for some $t$ we have $E_t = \emptyset$, consider the moment all
nodes have terminated. Then for all $v$ we have $N_v = \emptyset$. 
By propostion~\ref{prop-N-E} the proposition follows.
\qed
\end{proof}

\begin{proposition}
\label{prop-locmax}
Matching edge $e_i$ is a locally heaviest edge in $E_{i-1}$.
\end{proposition}
\begin{proof}
Let $e_i=(u,v)$.
By proposition~\ref{prop-c-E} $e_i \in E_{i-1}$.
To see that this is also the locally heaviest edge in $E_{i-1}$, suppose
an edge $(u,w) \in E_{i-1}$ is heavier. Then $w \in N_u$ by
proposition~\ref{prop-N-E}, but then $c_u = w$ instead.
\qed
\end{proof}

\begin{theorem}
Protocol~\ref{prot-dist} computes for any graph $G=(V,E)$ a 
matching $\matching{G}$ such that 
$w(\matching{G}) \le \frac{1}{2}w(\maxmatching{G})$ in time $O(|E|)$.
\end{theorem}
\begin{proof}
We first show that if protocol~\ref{prot-dist} computes a matching
$\matching{G}$, then there is a run of protocol~\ref{prot-seq} that returns the
same matching. Consider a run of protocol~\ref{prot-dist} on input $G$.
Let $x_i$ be the ordered sequence of matching events
in that run as defined above.

Now consider the sequential algorithm~\ref{prot-seq}. Define $E'_0 = E$, and
let $E'_i$ be the set of remaining edges in the graph after adding the $i$-th
edge $e'_i$ to the matching and removing the incident edges. Clearly
$E'_0 = E_0$. A simple inductive argument shows that $E'_i =  E_i$ for all $i$,
if we let protocol~\ref{prot-seq} select edge $e'_i = e_i$ 
(by proposition~\ref{prop-locmax} and the induction hypothesis
this is a locally heaviest edge and therefore a possible selection).

We conclude that the sequential algorithm adds the same edges to the matching
as the distributed algorithm in this run. According to proposition~\ref{prop-term},
for some $t$ we have $E_t = \emptyset$. Then also $E'_t = \emptyset$
so the sequential algorithm doesn't add any more edges. 
The bound on the approximation follows from
Lemma~\ref{lem-seqapprox}. The time complexity follows from
proposition~\ref{prop-onemess}.
\qed
\end{proof}

\section{Conclusions}

We have described a distributed algorithm that computes a $\frac{1}{2}$
approximation of the maximum matching of a weighted graph in $O(|E|)$ 
time, based on a sequential algorithm achieving the same approximation.

Other sequential algorithms, that improve the approximation to $3/2+\epsilon$ 
are known~\cite{drake2003maxmatching2,pettie2004maxmatching}. It is an open
question whether these algorithms can also be distributed, and if so, at which
cost in terms of running time.

\bibliography{/home/hoepman/lit/strings,weighted-matchings,/home/hoepman/lit/conferences}

\begin{thebibliography}{DH03b}

\bibitem[Avi83]{avis1983weightedmatching}
{\sc Avis, D.}
\newblock A survey of heuristics for the weighted matching problem.
\newblock {\em Networks {\bf 13}\/} (1983), 475--493.

\bibitem[CHS02]{chattopadhyay2002selfstabmatching}
{\sc Chattopadhyay, S., Higham, L., and Seyffarth, K.}
\newblock Dynamic and self-stabilizing distributed matching.
\newblock In {\em 21st \bibselect{PODC} {Ann.\ Symp.\ on Principles of
  Distributed Computing}\/} (Monterey, CA, USA, 2002), ACM Press, pp.~290--297.

\bibitem[DH03a]{drake2003maxmatching2}
{\sc Drake, D., and Hougardy, S.}
\newblock Improved linear time approximation algrotihms for weighted matchings.
\newblock In {\em 2764 7th Int.\ Workshop on Randomization and Approximation
  Techniques in Computer Science (APPROX)\/} (2003), no.~2764 in
  \bibselect{LNCS} {Lect.\ Not.\ Comp.\ Sci.\ }, pp.~14--23.

\bibitem[DH03b]{drake2003weightedmatching}
{\sc Drake, D., and Hougardy, S.}
\newblock A simple approximation algorithm for the weighted matching problem.
\newblock {\em \bibselect{Inf.\ Proc.\ Letters} {Information Processing
  Letters} {\bf 85}\/} (2003), 211--213.

\bibitem[Gab90]{gabow1990weightedmatching}
{\sc Gabow, H.}
\newblock Data structures for weighted matching and nearest common ancestors
  with linking.
\newblock In {\em 1th \bibselect{SODA} {Symposium on Discrete Algorithms}\/}
  (San Fransisco, Ca., USA, 1990), ACM, pp.~434--443.

\bibitem[II86]{israeli1986maxmatching}
{\sc Israeli, A., and Itai, A.}
\newblock A fast and simple randomized parallel algorithm for maximal matching.
\newblock {\em \bibselect{Inf.\ Proc.\ Letters} {Information Processing
  Letters} {\bf 22}\/} (1986), 77--80.

\bibitem[KS00]{karaata2000selfstabmatching}
{\sc Karaata, M., and Saleh, K.}
\newblock A distributed self-stabilizing algorithm for finding maximal
  matching.
\newblock {\em Computer Systems Science and Engineering {\bf 3}\/} (2000),
  175--180.

\bibitem[MV80]{micali1980maxmatching}
{\sc Micali, S., and Vazirani, V.}
\newblock An {$O(\sqrt{V}E)$} algorithm for finding maximum matching in general
  graphs.
\newblock In {\em 21nd \bibselect{FOCS} {Symp.\ on Foundations of Computer
  Science}\/} (??, 1980), IEEE Comp. Soc. Press, pp.~17--27.

\bibitem[PS04]{pettie2004maxmatching}
{\sc Pettie, S., and Sanders, P.}
\newblock A simple linear time {$2/3 - \epsilon$} approximation for maximum
  weight matching.
\newblock Tech. Rep. MPI-I-2004-1-002, Max-Planck-Institut f{\"{u}}r
  Informatik, Saarbr{\"{u}}cken, germany, 2004.

\bibitem[Pre99]{preis1999maxmatching}
{\sc Preis, R.}
\newblock Linear time 1/2-approximation algorithm for maximum weighted matching
  in general graphs.
\newblock In {\em 16th \bibselect{STACS} {Ann.\ Symp.\ on Theoretical Aspects
  of Computer Science}\/} (Trier, Germany, 1999), C.~Meinel and S.~Tison
  (Eds.), \bibselect{LNCS} {Lect.\ Not.\ Comp.\ Sci.\ } 1563, Springer,
  pp.~259--269.

\bibitem[UC00]{uehara2000maxmatching}
{\sc Uehara, R., and Chen, Z.}
\newblock Parallel approximation algorithms for maximum weighted matching in
  general graphs.
\newblock {\em \bibselect{Inf.\ Proc.\ Letters} {Information Processing
  Letters} {\bf 76}\/} (2000), 13--17.

\bibitem[WW04]{wattenhofer2004weightedmatching}
{\sc Wattenhofer, M., and Wattenhofer, R.}
\newblock Distributed weighted matching.
\newblock In {\em 18th \bibselect{DISC} {Int.\ Symp.\ Distrbuted Computing}\/}
  (Amsterdam, the Netherlands, 2004), R.~Guerraoui (Ed.), \bibselect{LNCS}
  {Lect.\ Not.\ Comp.\ Sci.\ } 3274, Springer, pp.~335--348.

\end{thebibliography}

\end{document}